\documentclass[12pt]{article}

\usepackage{amsmath,amssymb}
\usepackage{cite}

\newcommand{\dd}{\partial}
\newcommand{\m}{\mu}
\newcommand{\n}{\nu}
\newcommand{\ls}{\left(}
\newcommand{\rs}{\right)}
\newcommand{\al}{\alpha}
\newcommand{\ff}{\varphi}
\newcommand{\te}{\theta}
\newcommand{\str}[1]{\mathrel{\mathop{\longrightarrow}\limits_{#1}}}
\newcommand{\de}{\delta}
\newcommand{\na}{\nabla}
\newcommand{\sh}{\sinh}
\newcommand{\ch}{\cosh}

\newcommand{\disn}[2]{$$\displaylines{\refstepcounter{equation}%
            \label{#1}\hskip 1em minus 1em #2\hfilneg}$$}
\newcommand{\nom}{\hfil\hskip 1em minus 1em (\theequation)}
\newcommand{\no}{\hfil \hskip 1em minus 1em\phantom{(\theequation)}%
            \hfilneg\cr\hfilneg\hskip 1em minus 1em\hfil}

\begin{document}

\title{Hawking into Unruh mapping for embeddings\\ of hyperbolic type}
\author{S.A.~Paston\thanks{E-mail: paston@pobox.spbu.ru}\\
{\it Saint Petersburg State University, Saint Petersburg, Russia}
}
\date{\vskip 15mm}
\maketitle

\begin{abstract}
We study the conditions of the existence of Hawking into Unruh mapping for hyperbolic (Fronsdal-type) embeddings of metric into the Minkowski space, for which timelines are hyperbolas.
Many examples are known for global embeddings into the Minkowskian spacetime (GEMS)
with such mapping for physically interesting metrics with some symmetry.
However the examples of embeddings, both smooth and hyperbolic, for which there is no mapping, were also given.
In the present work we prove that
Hawking into Unruh mapping takes place for a hyperbolic embedding of an arbitrary metric with a time-like Killing vector and a Killing horizon
if the embedding of such type exists and smoothly covers the horizon.
At the same time we do not assume any symmetry (spherical for example), except the time translational invariance which corresponds to the existence of a time-like Killing vector.
We show that the known examples of the absence of mapping do not satisfy the formulated conditions of its existence.
\end{abstract}

\newpage

\section{Introduction}
The Hawking effect \cite{hawking75} predicts that in a spacetime with an event horizon
the observer situated outside the horizon will see a flow of particles with thermal distribution.
The nature of this quantum effect is in a non trivial character of the causal structure of the spacetime with the horizon in which the quantum field theory is considered.
Due to this non triviality the vector of state being vacuum from the point of view of field operators at the initial surface,
being decomposed over the basis of states corresponding to field operators at the finite surface in the future
(which includes the infinitely distant surface and the event horizon),
corresponds to a thermal radiation.

The temperature of this radiation measured by the observer (the Hawking temperature) is given by the formula
\disn{1}{
T_H=\frac{k}{2\pi}
\nom}
via the so called surface gravity $k$,
defined for a given observer as follows (see for example \cite{townsend}).
Let $\xi$ be a time-like Killing vector normalized in the manner that its value $\xi_0$ in the point where the observer stays satisfies the condition $\xi_0^2=1$.
Let the spacetime contain a null hypersurface $\cal N$,
at which $\xi$ is orthogonal to all vectors tangent to $\cal N$
(in this case the vector $\xi$ itself is a light-like vector tangent to $\cal N$),
then $\cal N$ is called the Killing horizon.
Then in the points of the surface $\cal N$ the following condition is satisfied
\disn{2}{
\xi^\al\na_\al\xi^\m=k\xi^\m,
\nom}
which defines the surface gravity  $k$ (here and below $\al,\m,\ldots=0,1,2,3$,
and $\na_\al$ means the covariant derivative).
From the physical point of view $k$ corresponds to a 4-dimensional acceleration
of the particle that is located near the horizon so as the Killing vector $\xi$
is tangent to its world line, at that the acceleration being rescaled taking
into account the red shift for an observer staying in some point outside the horizon.

For the 3-dimensionally infinite spaces for which $\xi^2$ is limited at the 3-dimensional infinity, for example for most black holes,
the observer is usually assumed to be at the infinity,
then the Killing vector is normalized by the condition $\xi^2\str{r\to\infty}1$.
However if we consider the spaces for which the 3-dimensional infinity is absent (like for the closed Friedmann-Robertson-Walker model), for example the de Sitter space, we must analyze the Hawking effect
from the point of view of the observer staying at a finite distance.

The Unruh effect \cite{unruh} predicts that while an inertial observer in a Minkowski space sees a vacuum state, an observer moving with a constant acceleration $w$ will see a thermal radiation with the Unruh temperature
\disn{2.1}{
T_U=\frac{w}{2\pi}.
\nom}
This effect is usually described using an Unruh-DeWitt detector moving over a corresponding trajectory and interacting with a quantum field existing in the spacetime.
A similar effect is also found for other accelerated motions of the detector, but in this case the radiation spectrum is no longer thermal \cite{letaw81}.

An observer moving with a constant acceleration is stationary in the Rindler coordinates \cite{rindler} $t,\rho,x^{2,3}$, related to the Lorenz coordinates $x^\m$ by the relation
\disn{3}{
x^0=\rho\sh t,\qquad x^1=\rho\ch t
\nom}
or, in the light-like coordinates $x^\pm=x^0\pm x^1$,
\disn{4}{
x^+=\rho e^t,\qquad x^-=-\rho e^{-t}.
\nom}
The time $t$ translation  for the Rindler coordinates corresponds to the Lorenz boost in the Minkowski space.
The corresponding Killing vector $\xi^\m$, normalized by the unity at $\rho=\rho_0$, has the components
\disn{5}{
\xi^+=\frac{1}{\rho_0}\rho e^t=\frac{1}{\rho_0}x^+,\qquad \xi^-=\frac{1}{\rho_0}\rho e^{-t}=-\frac{1}{\rho_0}x^-,
\qquad \xi^{2,3}=0.
\nom}
For such Killing vector the Killing horizon $\cal N$ exists
being the surface $(x^+=0\bigcup x^-=0)\bigcap x^1\ge 0$,  therefore we can define the surface gravity.

Taking into account that in Lorenz coordinates in the Minkowski space the covariant derivative $\na_\al$ reduces to the conventional one $\dd_\al$,
one can easily calculate the left side of the equation (\ref{2}) in some point of the horizon, for example in which $x^-=0,x^+>0$:
\disn{6}{
\xi^\al\na_\al\xi^\m=\xi^+\dd_+\xi^\m=\de^\m_+\frac{x^+}{\rho_0^2}=\frac{1}{\rho_0}\xi^\m.
\nom}
That gives us the surface gravity value from the point of view of the observer with the coordinate $\rho_0$:
\disn{7}{
k=\frac{1}{\rho_0}.
\nom}
Clearly, this value is equal to the constant acceleration $w$ of this observer,
hence one can say that for the Rindler observer in the Minkowsky space the Hawking temperature (\ref{1}) and the Unruh temperature (\ref{2.1}) are equal,
i.e. the most simple case of Hawking into Unruh mapping takes place.

More complex cases of such mapping also named GEMS (global embedding Minkowskian spacetime) approach
were found in \cite{deserlev98,deserlev99} at the consideration of isometric embeddings of the Schwarzschild metric, de Sitter metric and some others in a flat ambient space.
For isometric embeddings the metric is expressed via the embedding function $y^a(x^\m)$ by an induced metric formula
\disn{7.1}{
g_{\m\n}=(\dd_\m y^a) (\dd_\n y^b) \eta_{ab},
\nom}
where $\eta_{ab}$ is the metric of a flat ambient space (here and below $a,b,\ldots=0,1,2,\ldots,N-1$, $N$ is the dimensionality of the ambient space).
It has been found out in \cite{deserlev98,deserlev99} that the temperature $T_H$ corresponding to the Hawking effect caused by the presence of the horizon is equal, for the considered embeddings, to the Unruh temperature $T_U$ caused by a uniformly accelerated motion of the observer from the point of view of the ambient space.
The motion appears to be uniformly accelerated due to the use of hyperbolic embeddings under the form
\disn{7.1a}{
y^0=\rho(x^1,x^2,x^3)\sh x^0,\qquad
y^1=\rho(x^1,x^2,x^3)\ch x^0,\qquad
y^A=y^A(x^1,x^2,x^3),
\nom}
where $A=2,\ldots,N-1$. For the embeddings of such type the timelines are hyperbolas.
Note that for the embeddings of such type we have
$g_{0i}=0$ (where $i=1,2,3$).

The obtained result caused a number of works where the Hawking into Unruh mapping was tested
for many other metrics.
The corresponding results were  obtained for various types of black holes (including those with an electric charge, with a cosmological constant and with the dimensionality other than 4), black strings and wormholes,
see \cite{kim00,lemos,arXiv:1012.5709,hep-th/0103036,gr-qc/0303059,arXiv:1311.0592} and cited herein.
A detector moving over a circumference \cite{gr-qc/0409107} or in a free fall \cite{arXiv:0805.1876} near the black hole was also discussed.
In this case the radiation spectrum is no more exactly thermal.
It has been shown in \cite{arXiv:0901.0466} that the coincidence is still valid for the corrections to the Hawking and Unruh temperatures.
It has been noted in \cite{Banerjee} that the Hawking into Unruh mapping takes place if we embed the $(t-r)$ sector of the Riemannian space only.
Note that in all cases the hyperbolic type embeddings were used where the timelines are hyperbolas in the ambient space
(such embeddings can also be called "Fronsdal-type embeddings"{}, because for the Schwarzschild metric an embedding of such type was suggested by C.~Fronsdal \cite{frons}).

GEMS approach allows to reduce the thermal properties of the space with horizon to the kinematical Unruh effect due to Rindler motion in ambient space, which can simplify their study. Besides, the
existence of the Hawking into Unruh mapping can mean a strong relation between quantum theories in a Riemannian space and in a corresponding flat ambient space.
Such relation supports the idea that gravity, usually described in the framework of General Relativities as a result of the Riemannian space curvature, can be formulated as an embedding theory where the 4-dimensional spacetime is a surface in a flat space with higher dimensionality.
Just in this flat space the quantum theory can be formulated, and such approach can help to construct a correct quantum gravitation theory, especially when
the gravity is formulated as a field theory in a flat ambient space where the matter fields are also described by some functions in the ambient space.
In the formulation of the gravity in the form of a field theory (such approach was proposed in \cite{statja25})
this field describes the splitting of ambient space into a system of 4-dimensional surfaces
and each of these surfaces can be considered as our spacetime.
Various modifications of the embedding theory were developed in \cite{regge,deser,pavsic85let,tapia,davkar,statja18,statja24,faddeev,statja26} and other works.

However we must note that Hawking into Unruh mapping takes place not for all embeddings of Riemannian spaces.
In \cite{statja34} it has been noted that there is no mapping for Schwarzschild and Reissner-Nordstr\"om metrics (suggested in \cite{statja27} and \cite{statja30}, respectively), smoothly covering the horizons and whose type is not hyperbolic.
The same work contains also other examples of embeddings for which there is no mapping:
non hyperbolic type embeddings -- for the de-Sitter metrics,
as well as hyperbolic type embeddings -- for the de Sitter and anti de Sitter metrics,
and also for the Minkowski metric.

All known embeddings for which mapping takes place are hyperbolic
(that seems natural due to the fact that the Unruh radiation spectrum in this case is for sure thermal),
however the examples of hyperbolic embeddings with no mapping are also known.
Hence the problem arise to formulate the most general conditions of the existence of mapping for hyperbolic embeddings
allowing to predict the existence of mapping for a particular case without direct verification.
The present article aims to solve this problem in the most general case.
In particular, we do not assume the existence of any symmetry (spherical for example), except the translational invariance with respect to time corresponding to the existence of a time-like Killing vector.
In Section~\ref{hyp} we discuss the general form of a hyperbolic embedding with horizon and we analyze its smoothness.
In Section~\ref{uslov} we prove the main statement on the existence of mapping,
and in Section~\ref{prim} we discuss the known examples of embeddings for which there is no mapping, and we show that their existence does not contradict the proven statement.

\section{General form of a hyperbolic embedding with horizon}\label{hyp}
Consider a 4-dimensional surface $\cal M$ being an hyperbolic embedding of
some
metric (which is not in general case a solution of Einstein equations with some fixed matter)
with a time-like Killing vector $\xi$.
It should be noted that the hyperbolic embedding exists not for each metric,
see about the construction of embeddings of various types in
\cite{statja27,statja30,statja32}.
For the existence of Hawking radiation we assume the existence of the Killing horizon
i.~e. a null hypersurface $\cal N$, on which $\xi$ is normal to $\cal N$ (see for example \cite{townsend}).
Then from the fact that $\cal N$ is null hypersurface it follows that
$\xi^2=0$ on $\cal N$ and $\xi$ is tangent to $\cal N$ on $\cal N$.
We choose the coordinates $t,\rho,x^2,x^3$ in order to make $\xi$ tangent to the time lines $t$,
and to make the coordinate $\rho$ equal to zero at $\cal N$, so that $\xi^2\str{\rho\to0}0$.
Let us assume that the observer (an Unruh-DeWitt detector for example) is stationary in a selected coordinate system and has constant coordinates $\rho=\rho_0>0,x^2,x^3$,
and also that $\xi^2>0$ at $0<\rho\le \rho_0$.

The existence of a time-like Killing vector means that
the metric is symmetrical with respect to the one-dimensional translation group of time shifts.
In order to build a surface $\cal M$ being also symmetric with respect to this group
we have to consider the representations of the translation group
whose matrices correspond to the transformations of the Poincar\'e group of the ambient space
(see the details of the method of constructing the symmetrical embeddings in \cite{statja27}).
In one of the most simple cases the matrices mentioned above correspond to Lorenz boosts in the ambient space, i.e. the shift $t$ is equivalent to the boost.
In this case the time lines in the ambient space appear to be hyperbolas
and the embeddings of such type are called hyperbolic
(or, as already mentioned above, the "Fronsdal-type embeddings"{}).
The arbitrariness in the choice of coordinates $\rho,x^2,x^3$ allows to write an arbitrary hyperbolic-type embedding in the region $0\le \rho\le \rho_0$ (where $\xi^2>0$) under the form
\disn{8}{
y^0=\rho\sh(\al t),\qquad
y^1=\rho\ch(\al t),\qquad
y^A=y^A(\rho,x^2,x^3).
\nom}
Here and below $A=2,\ldots,N-1$, and $\al$ is an arbitrary positive constant introduced for generality.
Let us stress
that the components $y^A$ of the embedding function $y^a(x^\m)$ depend
in some way on the coordinates $\rho,x^2,x^3$, but not of $t$.

Note that hyperbolic embedding (\ref{8}) in chosen coordinates has a property $g_{0i}=0$ (where $i=1,2,3$).
For the metrics of general form for which the time-like Killing vector exists and the time is chosen in such a way that it is tangent to the time lines, the condition $g_{0i}=0$ can be fulfilled not always.
In this case the embeddings of hyperbolic type for such metrics do not exist.

Let us normalize the Killing vector $\xi$ corresponding to the time shift $t$ in order to make it a unit vector in the observation point $\rho=\rho_0$.
Then in the coordinates $t,\rho,x^2,x^3$ its components will read
\disn{9}{
\xi^\m=\de^\m_0\frac{1}{\sqrt{g_{00}(\rho_0)}}=\de^\m_0\frac{1}{\al \rho_0},
\nom}
and in the ambient space it will have a corresponding vector $\xi^a=\xi^\m\dd_\m y^a$ with the components
\disn{10}{
\xi^0=\frac{\rho}{\rho_0}\ch(\al t),\qquad
\xi^1=\frac{\rho}{\rho_0}\sh(\al t),\qquad
\xi^A=0.
\nom}

Suppose that the embedding smoothly covers the horizon, i.e. the considered four-dimensio\-nal surface is smooth at $\rho=0$.
Since the coordinate $t$ appears to be singular at the horizon (as for example the time in the Schwarzschild coordinates for the black hole),
the embedding function for this coordinate choice needs not to be explicitly smooth,
so its smoothness cannot be verified directly.
In order to choose nonsingular coordinates it is useful to prove that the plane $y^0,y^1$ is tangent to $\cal M$ at the $y^0=y^1=0$ point. For this proof let us note that the points with any fixed value of $t$ tend to this point in the limit $\rho\to0$ according to (\ref{8}). At $\rho>0$ the vector $\xi^a/\rho$ is tangent to the surface $\cal M$, therefore if we use (\ref{10}) and consider this vector in the limit $\rho\to0$ at the different values of $t$, then we can conclude that  every vector lying in the plane $y^0,y^1$ will be tangent to $\cal M$ in the point $y^0=y^1=0$ (provided that $\cal M$ is smooth). It means that the plane $y^0,y^1$ is tangent to $\cal M$ in this point.
Hence the values $y^0,y^1$ can be used as coordinates at $\cal M$ in the vicinity of the point $y^0=y^1=0$,
and the smoothness of $\cal M$ ensures the smoothness of the embedding function written in these coordinates.
If we use the values $y^0,y^1,x^2,x^3$ in the given vicinity as the coordinates for $\cal M$,
the embedding function (\ref{8}) (in the region where this formula is applicable, i.e. at ${y^1}^2-{y^0}^2\ge0$) takes the form
\disn{11}{
y^0=y^0,\qquad
y^1=y^1,\qquad
y^A=y^A\ls\sqrt{{y^1}^2-{y^0}^2},x^2,x^3\rs.
\nom}
Its smoothness with respect to $y^0,y^1$ means a smooth dependence of the function $y^A(\rho,x^2,x^3)$ on ${\rho}^2$.
Therefore
in the vicinity of the horizon that corresponds to the points of the ambient space satisfying  the equation
\disn{11.1}{
y{^1}^2-{y^0}^2=0,
\nom}
the following expansion is fulfilled:
\disn{12}{
y^A=f^A(x^2,x^3)+\ls{y^1}^2-{y^0}^2\rs h^A(x^2,x^3)+O\ls\ls{y^1}^2-{y^0}^2\rs^2\rs.
\nom}
This expansion being satisfied is hence a criterion whether the considered embedding smoothly covers the horizon.
This criterion will be used below in Section~\ref{prim}.

\section{Conditions of the existence of mapping}\label{uslov}
Let us study the conditions of the existence of Hawking into Unruh mapping for an embedding of the form (\ref{8}) smoothly covering the horizon.
Since we consider the hyperbolic embeddings we limit ourself by the consideration of the metrics
for which such embedding exists.
In order to define the temperature $T_H$ of the Hawking radiation caused by the presence of the horizon,
we find out from (\ref{2}) the value of the surface gravity $k$ from the point of view of the observer with the coordinate $\rho=\rho_0$.
To do that we rewrite this equation in terms of the vector $\xi^a$ of the ambient space (\ref{10}).
The obtained equation will be equivalent to the initial one if the embedding smoothly covers the horizon
(an example of a nonsmooth embedding for which the equivalence is lost is considered below in the Section~\ref{prim}).
The covariant derivative of the vector in the framework of the embedding theory formalism can be written under the form of (this formalism is exposed in~\cite{statja18})
\disn{13}{
\na_\al\xi^\m=e^\m_a\dd_\al\ls\xi^\n e_\n^a\rs,
\nom}
where
\disn{14}{
e_\n^a=\dd_\n y^a,\qquad
e^\m_a=g^{\m\n}e_\n^b \eta_{ab},
\nom}
and $\eta_{ab}$ is a flat metric of the ambient space. Multiplying the equation (\ref{2}) by $e_\m^b$ and using the formula (\ref{13}), we get the equation
\disn{15}{
\Pi^b_a\xi^\al\dd_\al\xi^a=k\xi^b,
\nom}
where the value
\disn{16}{
\Pi^b_a=e^b_\n e^\n_a
\nom}
is the projector on the space tangent to the surface $\cal M$ at the given point.

For the convenience of the analysis of the equation (\ref{15}) we introduce the light-like coordinates $y^\pm=y^0\pm y^1$ in the ambient space,
similarly to the case considered above (see formulas (\ref{3})-(\ref{5})) in the Unruh effect discussion.
In these coordinates the formula (\ref{8}) takes the form
\disn{17}{
y^+=\rho e^{\al t},\qquad
y^-=-\rho e^{-\al t},\qquad
y^A=y^A(\rho,x^2,x^3),
\nom}
allowing to rewrite (\ref{10}) as
\disn{18}{
\xi^+=\frac{\rho}{\rho_0}e^{\al t}=\frac{y^+}{\rho_0},\qquad
\xi^-=\frac{\rho}{\rho_0}e^{-\al t}=-\frac{y^-}{\rho_0}.\qquad
\xi^A=0.
\nom}

It has been shown in Section~\ref{hyp} that the embedding function of the surface $\cal M$ is smooth in the coordinates $y^0,y^1,x^2,x^3$ (see~(\ref{11})),
hence it will also be smooth in the coordinates $y^+,y^-,x^2,x^3$, where it takes the form
\disn{19}{
y^+=y^+,\qquad
y^-=y^-,\qquad
y^A=y^A\ls\sqrt{y^+ y^-},x^2,x^3\rs.
\nom}
We write the equation (\ref{15}) in these coordinates in the point of the horizon for which $y^-=0,y^+>0$ (that corresponds to the horizon equation (\ref{11.1})). Note that in this point $\xi^-=0$ due to (\ref{18}).
For the left hand side of the equation (\ref{15}) we obtain:
\disn{20}{
\Pi^b_a\xi^\al\dd_\al\xi^a=\Pi^b_a\xi^+\dd_+\xi^a=
\frac{y^+}{\rho_0^2}\Pi^b_a m^a,
\nom}
where $m^a$ is a constant vector with the components $m^+=1;\,m^-=0;\,m^A=0$.
It is easy to note from (\ref{19}) that at $y^-=0$ the vector $m^a$ is tangent to $\cal M$, because it coincides with $\dd_+ y^a$. Hence, following the projector properties, we obtain $\Pi^b_a m^a=m^b$,
and finally the equation (\ref{15}) in the considered point can be rewritten under the form
\disn{21}{
\frac{y^+}{\rho_0^2}m^b=k\frac{y^+}{\rho_0}m^b.
\nom}
We can easily found from it the value of the surface gravity for the observer with the coordinate $\rho=\rho_0$:
\disn{22}{
k=\frac{1}{\rho_0}.
\nom}

It means that, according to the formula (\ref{1}), the observer will see the Hawking radiation with the temperature $T_H=1/(2\pi\rho_0)$.
On the other hand, as seen from (\ref{8}), this observer in the ambient space moves with a constant acceleration $w=1/\rho_0$,
which corresponds, according to the formula (\ref{2.1}), to the Unruh temperature $T_U$, which appears to be equal to $T_H$.
Hence we have proven the following statement.
\vskip 0.5em

\indent\textit{\textbf{Statement:} Consider an arbitrary metric with a time-like Killing vector
and a Killing horizon,
for which hyperbolic embedding exist.
If a hyperbolic embedding of this metric (which can always be written under the form (\ref{8})) smoothly covers the horizon, then the Hawking into Unruh mapping takes place.}
\vskip 0.5em

One should note that the reasoning performed during the proving of this statement
mostly reproduces the reasoning used above at the discussion of the Unruh effect.
The only difference is that during the deduction of the formula (\ref{22})
we had to take into account the non trivial nature of the covariant derivative,
while at the deduction of (\ref{7}) the covariant derivative is equal to the conventional one.
However in both cases the results are the same -- Hawking into Unruh mapping takes place -- and this can be explained as follows.

Note that the difference between the generic form of the hyperbolic embedding (\ref{8}) and the form of the Minkowski space in the Rindler coordinates (\ref{3})
is only that for the Minkowski space the "transverse"{} components $x^{2,3}$ do not depend on $\rho$,
while for the hyperbolic embedding the analogous "transverse"{} components $y^A$ depend on $\rho$.
However this dependence at small $\rho$ for smooth embeddings has no term linear over $\rho$ (see the expansion (\ref{12})).
Hence in a small vicinity of the horizon the smooth surface $\cal M$ at fixed $x^{2,3}$ coincides at first approximation with the Minkowski space,
so for the observer whose coordinate $\rho$ is small the Hawking and Unruh temperatures are trivially equal.
For the observers with arbitrary values of the coordinate $\rho$ this coincidence remains valid automatically
(this has already been mentioned in \cite{statja34}) due to the relation following from (\ref{2.1}) and (\ref{8}):
\disn{23}{
T_U\sqrt{g_{00}}=\frac{1}{2\pi\rho}\sqrt{\al^2\rho^2}=const
\nom}
and due to the well known Tolman law having the same form
\disn{24}{
T_H\sqrt{g_{00}}=const.
\nom}
This reasoning allows to easily explain the result obtained in the present work.

Note that the statement about the existence of mapping was proven under the  assumptions being rather general.
Concerning the spacetime metric we assume only the existence
of a time-like Killing vector and a Killing horizon (without this assumption one can not
consider the Hawking radiation) and suppose the  existence of hyperbolic embedding.
The embedding function was taken in its most general form, corresponding to the hyperbolic type
of the implementation of the translational invariance with respect to time shifts.
The presence of any other symmetry, spherical for example, was not assumed specially.

\section{The analysis of the examples of mapping absence}\label{prim}
It has been noted in \cite{statja34} that Hawking into Unruh mapping takes place not for all embeddings.
We will analyze the examples given there and verify their correlation with the statement proven in the previous section.

First of all, it has been noted that mapping does not work for non hyperbolic embeddings even if they smoothly cover the horizon.
The examples of such embeddings are the Schwarzschild and Reissner-Nordstr\"om metrics embeddings suggested in \cite{statja27} and  \cite{statja30}, respectively.
The non hyperbolic embeddings can also be easily obtained by a trivial isometric bending (such as bending of a plane into a part of a cylinder) of a flat embedding space which contains a surface for which, in particular, mapping existed before bending.
The example of such bending for the de Sitter space was given in \cite{statja34}.
Since in the discussed statement only hyperbolic embeddings are mentioned, the given examples do not contradict it.

Then an example was given of a hyperbolic type embedding, which is obtained by a corresponding flexion of a flat space. It has the form
 \disn{25}{
\begin{array}{cl}
&y^0=\al^{-1} \sh (\al t),\no
&y^1=\al^{-1} \ch (\al t),\no
&y^2=r\cos\te,\no
&y^3=r\sin\te\cos\ff,\no
&y^4=r\sin\te\sin\ff
\end{array}
\nom}
with the ambient space signature $(+----)$.
There is no Hawking effect for this embedding, because its metric coincides with the Minkowski space metrics,
while the Unruh effect with the temperature $T_U=\al/(2\pi)$ takes place.
The given example does not comply with the conditions of the statement under discussion because of the absence of the horizon.
Note that for this reason the embedding (\ref{25}) does not at all correspond to the general form (\ref{8}) for which the existence of a horizon was assumed.

There was also an example of a hyperbolic embedding of the de Sitter space whose interval can be written under the form
 \disn{25.1}{
ds^2=R^2\cos^2\!\chi\, dt^2-R^2 d\chi^2-R^2\sin^2\!\chi\, d\te^2-R^2\sin^2\!\chi\,\sin^2\!\te\, d\ff^2.
\nom}
This embedding has the form
 \disn{26}{
\begin{array}{cl}
&y^0=R\al^{-1}\cos\chi\,\sh(\al t),\no
&y^1=R\al^{-1}\cos\chi\,\ch(\al t),\no
&y^2=R\, \sin\chi\,\cos\te,\no
&y^3=R\, \sin\chi\,\sin\te\,\cos\ff,\no
&y^4=R\, \sin\chi\,\sin\te\,\sin\ff,\no
&y^5=R\sqrt{1-\al^{-2}}\cos\chi
\end{array}
\nom}
with the ambient space signature $(+-----)$.
Hawking into Unruh mapping works in this case only at $\al=1$, when this space transforms into a standard de Sitter hyperboloid. At $\al>1$ there is no mapping,
because the Hawking temperature $T_H$ depends only on the internal geometry (as seen from (\ref{1}),(\ref{2})) and hence does not depend on $\al$,
while the Unruh temperature, as follows from (\ref{2.1}),(\ref{26}), is
\disn{27}{
T_U=\frac{\al}{2\pi R\cos\chi}.
\nom}

The embedding (\ref{26}) is hyperbolic (it corresponds to the form (\ref{8}) at $\rho=R\al^{-1}\cos\chi$),
and in the used coordinates it contains the horizon at $\chi=\pi/2$.
This is why its existence does not contradict the statement formulated in Section~\ref{uslov}
only if the last condition of the assumption, the smoothness at the horizon, is violated.
The smoothness is indeed absent according to the criterion (\ref{12}),
because when the embedding (\ref{26}) is written under the form (\ref{11}), one of the components of the embedding function has the form
 \disn{28}{
y^5=\sqrt{\al^2-1}\sqrt{{y^1}^2-{y^0}^2},
\nom}
which does not correspond to the expansion (\ref{12}) if $\al>1$.

Now we will study in details the structure of the singularity of the embedding (\ref{26}) for $\al>1$ at the horizon to which corresponds the set of points of the ambient space
 \disn{29}{
\ls y^0=y^1\bigcup y^0=-y^1\rs\bigcap y^1\ge0 \bigcap y^5=0\bigcap {y^2}^2+{y^3}^2+{y^4}^2=R^2.
\nom}
We note for this purpose that the corresponding surface can be defined by two equations for the coordinates of the ambient space:
 \disn{30}{
{y^0}^2-{y^1}^2-{y^2}^2-{y^3}^2-{y^4}^2-{y^5}^2=-R^2,
\nom}\vskip -2em
 \disn{31}{
{y^1}^2={y^0}^2+\frac{{y^5}^2}{\al^2-1}.
\nom}
The second of these equations being the cone equation.
The equation (\ref{30}) does not forbid the vicinity of cone vertex to belong to the surface,
hence the surface  at $\al>1$ in some of the points corresponding to the horizon, namely at $y^0=y^1=y^5=0$, has a conic singularity.

The existence of a conic singularity at the surface given by (\ref{26}) leads to the fact that
the value of the surface gravity $k$ calculated by the method given in Section~\ref{uslov} which uses the expression for the embedding function
begins to depend on its parameter $\al$ which is not contained in the metric.
This has the following reason:
if the surface is smooth, then the change of $\al$ corresponds to   an isometric bending.
But in our case this change appears to be a singular transformation
at which the points with finite values of the embedding functions tend to the infinity or to a singular point
(this can be seen from the analysis of the limit $\rho\to0,t\to\infty$ for the embedding function (\ref{17})).
As a result, when there is a singularity (at $\al>1$ in our case)
the value of $k$ defined by this way is not inevitably a value of a surface gravity from the point of view of the internal space-time geometry.
Consequently, the Hawking temperature $T_H$ also takes an incorrect value (depending on $\al$).
The considered example demonstrates the importance of the smoothness condition in the statement proven in Section~\ref{uslov}.

In \cite{statja34} a similar example of the absence of Hawking into Unruh mapping is also mentioned.
This is the anti de Sitter space embedding suggested in \cite{willison1302} for which
 \disn{31.1}{
ds^2=R^2\ch^2\!\chi\, dt^2-R^2 d\chi^2-R^2\sh^2\!\chi\, d\te^2-R^2\sh^2\!\chi\,\sin^2\!\te\, d\ff^2.
\nom}
This embedding can be written under the form (in \cite{willison1302} it has been constructed for an arbitrary dimensionality)
 \disn{32}{
\begin{array}{cl}
&y^0=R\al^{-1}\ch\chi\,\sh(\al t),\no
&y^1=R\al^{-1}\ch\chi\,\ch(\al t),\no
&y^2=R\, \sh\chi\,\cos\te,\no
&y^3=R\, \sh\chi\,\sin\te\,\cos\ff,\no
&y^4=R\, \sh\chi\,\sin\te\,\sin\ff,\no
&y^5=R\sqrt{1+\al^{-2}}\ch\chi
\end{array}
\nom}
with the ambient space signature $(+----+)$.
Its Unruh temperature $T_U$, as well as for the embedding (\ref{26}), depends on the arbitrary parameter $\al$ which is not contained in the metric.

Although the surface corresponding to this embedding can be defined, similarly to (\ref{30}), by two equations for the coordinates of the ambient space
 \disn{33}{
{y^0}^2-{y^1}^2-{y^2}^2-{y^3}^2-{y^4}^2+{y^5}^2=R^2,
\nom}\vskip -2em
 \disn{34}{
{y^1}^2={y^0}^2+\frac{{y^5}^2}{\al^2+1},
\nom}
the last of which being the cone equation,
the surface corresponding to (\ref{32}) has no singularity, because the equation (\ref{33}) does not allow the value $y^0=y^1=y^5=0$.
Hence (\ref{32}) is a smooth hyperbolic embedding.
It is however easy to note that it does not contain the horizon, because, if we transform it to the form (\ref{8}), we obtain $\rho=R\al^{-1}\ch\chi$, and $\rho$ cannot become zero.
This is why  the existence of this embedding, just as all other considered examples, does not contradict the statement proven in Section~\ref{uslov}.

In a result
we obtain that for hyperbolic embeddings smoothly covering the horizon the Hawking into Unruh mapping works,
and the GEMS method of the analysis of thermodynamic properties of spaces with horizons, suggested in \cite{deserlev98,deserlev99}, must always give good results.

\vskip 0.5em
\textbf{Acknowledgments}.
The author is grateful to A.A.~Sheykin for the useful discussions.
The work was partially supported by the Saint Petersburg State University grant N~11.38.660.2013.


\end{document}